\documentclass[superscriptaddress,twocolumn,showpacs,prl]{revtex4}

\usepackage{graphicx}
\usepackage{tabularx}
\usepackage{epsfig}
\usepackage{dcolumn}
\usepackage{bm}
\usepackage{amsmath}
\usepackage{amssymb}

\newcommand{\be}{\begin{equation}}
\newcommand{\ee}{\end{equation}}
\newcommand{\bea}{\begin{eqnarray}}
\newcommand{\eea}{\end{eqnarray}}

\begin{document}

\title{Temperature and Disorder Chaos in three-dimensional Ising Spin Glasses}

\author{Helmut G.~Katzgraber}
\affiliation{Theoretische Physik, ETH Z\"urich, 
CH-8093 Z\"urich, Switzerland}

\author{Florent Krz\c{a}ka{\l}a}
\affiliation{Laboratoire P.C.T., UMR CNRS 7083, ESPCI, 
10 rue Vauquelin, 75005 Paris, France}

\date{\today}

\begin{abstract}
We study the effects of small temperature as well as disorder
perturbations on the equilibrium state of three-dimensional Ising
spin glasses via an alternate scaling ansatz. By using Monte Carlo
simulations, we show that temperature and disorder perturbations
yield chaotic changes in the equilibrium state and that temperature
chaos is considerably harder to observe than disorder chaos.
\end{abstract}

\maketitle

The fragility of the equilibrium state of random frustrated
systems such as the Edwards-Anderson Ising spin glass
\cite{edwards:75,mezard:87,young:98} has been predicted a long time
ago \cite{mckay:82,parisi:84} and analyzed on the basis of scaling
arguments \cite{fisher:86,bray:87}. These scaling arguments predict
that the configurations which dominate the partition function change
{\em drastically} and {\em randomly} when the temperature or the
disorder in the interactions between the spins are modified ever
so slightly.  The {\em temperature chaos} and {\em disorder chaos}
effects have attracted considerable attention both from theory and
experiment because of their potential relevance in explaining the
spectacular rejuvenation and memory effects observed in hysteresis
experiments in spin-glasses \cite{nordblad:98,dupuis-ea:01,jonsson-ea:04} as
well as other materials, such as random polymers and pinned elastic
manifolds.  Although there is evidence of {\em disorder chaos}
in spin glasses, temperature chaos remains a controversial issue
\cite{kondor:89,neynifle:97,neynifle:98,billoire:00,billoire:02},
whereas for random polymers or pinned elastic objects
\cite{fisher:91d} chaos in general is well established
\cite{sales:02,silveira:04,doussal:05}.

Despite this lack of consensus, it has been surmised that temperature
chaos would only be observable in spin glasses at very large system
sizes and for large temperature changes \cite{aspelmeier:02a,rizzo:03}
thus making its presence unfathomable in simulations. These claims
have recently been challenged. In particular, recent results point
towards the existence of temperature chaos in four-dimensional Ising
spin glasses \cite{sasaki:05} where the free energy of a domain wall
induced by a change in boundary conditions changes its sign chaotically
with temperature in accordance with the droplet/scaling theories
\cite{mckay:82,fisher:86,bray:87}.  In this work we study the overlap
between states at different temperatures and disorder distributions
{\em directly} for a {\em physically relevant} three-dimensional (3D)
Ising spin glass. Our results show that the scaling laws that arise
from the droplet theory are indeed well satisfied in 3D provided low
enough temperatures are considered, although small corrections need
to be applied.  In addition, we show that temperature and disorder
chaos have similar scaling functions. By rescaling the characteristic
length scale in the problem, we show that disorder chaos appears at
much shorter scales than temperature chaos

The paper is organized as follows: We discuss first the model and Monte
Carlo methods used, followed by the disorder- and temperature-chaos
scaling approaches. We conclude with the results of our simulations
of the 3D Ising spin glass and a general discussion.

\paragraph*{Model and numerical method}

The Edwards-Anderson \cite{edwards:75} Ising spin glass is given by the
Hamiltonian
\be 
{\mathcal{H}}=-\sum_{\langle ij\rangle} J_{ij} S_i S_j ,
\label{Hamiltonian}
\ee
where the Ising spins $S_i \in \{\pm1\}$ are on a cubic lattice
with $N = L^3$ vertices and the interactions $J_{ij}$ are
Gaussian distributed random numbers with zero mean and standard
deviation unity. The sum is over nearest neighbor pairs. The
model undergoes a spin-glass transition at $T_{\rm c} = 0.951(9)$
\cite{bhatt:88,marinari:98,katzgraber:06}.

The order parameter of the system is defined via the overlap between two
copies $\alpha$ and $\beta$, i.e., 
\be
q_{\alpha,\beta}=\frac{1}{N} \sum_{i=1}^{N} S^{\alpha}_i S^{\beta}_i .
\label{olap}
\ee
Following previous studies~\cite{neynifle:97,neynifle:98}, we probe
temperature chaos when the temperature between both replicas
is shifted by an amount $\Delta T$. Disorder chaos is studied by 
introducing a perturbation $\Delta J$ in the
disorder, i.e., 
\be 
J_{ij} \rightarrow \tilde{J}_{ij} = \frac{J_{ij} + x \Delta J}{\sqrt{1+\Delta J^2}} 
\label{disshift}
\ee 
which leaves the disorder distribution invariant. In Eq.~(\ref{disshift})
$x$ is a Gaussian distributed random number with zero mean and standard
deviation unity.  To monitor the changes induced by the perturbations of the
system, we compute the {\em chaoticity parameter} 
\cite{neynifle:97,neynifle:98} given by 
\be Q_{T_1,T_2} = \left[
  \frac{ \langle q^2_{T_1,T_2}\rangle } {\sqrt{ \langle q^2_{T_1,T_1}\rangle
      \langle q^2_{T_2,T_2}\rangle }}\right]_{\rm av}
\label{parameter1}
\ee
for temperature chaos and by
\be
Q_{\Delta J} =  \left[\frac{ \langle q^2_{J_{ij},\tilde{J}_{ij}}\rangle 
                }{\langle q^2_{J_{ij},J_{ij}}\rangle}\right]_{\rm av} 
\label{parameter2}
\ee 
for disorder chaos, respectively. 

\newpage

In Eqs.~(\ref{parameter1}) and (\ref{parameter2}) $q^2$ is
the square overlap, Eq.~(\ref{olap}), between two copies at
different temperature/disorder.  Here $\langle \cdots \rangle$
represents a thermal average and $[\cdots]_{\rm av}$ represents
a disorder average.  In order to access low temperatures
necessary to probe temperature chaos, we have used the
parallel tempering \cite{hukushima:96,marinari:96} Monte Carlo
method in combination with the equilibration test presented in
Ref.~\cite{katzgraber:01}. Simulation parameters are listed in Table
\ref{tab:simparams}.

\begin{table}
\caption{Parameters of the simulation for each system size $L$. 
$N_{\rm samp}$ is the number of samples and $N_{\rm sw}$ is the total number 
of Monte Carlo sweeps used for equilibration for each of the $2N_T$ 
parallel tempering replicas for a single sample. An equal number of 
sweeps is used for measurement. The minimum temperature simulated is 
$T_{\rm min} = 0.20$, the highest $T_{\rm max} = 2.0$. For disorder chaos the
disorder shifts used in Eq.~(\ref{disshift}) are 
$\Delta J = 0.001$, $0.005$, $0.02$, $0.05$, $0.1$, $0.2$, and $0.5$. For
temperature chaos we compute the overlap between $T_{\rm min}$ and $T_i$ with
$i \in \{2,\ldots,N_{T}\}$ \cite{temperatures}.
\label{tab:simparams}}
\begin{tabular*}{\columnwidth}{@{\extracolsep{\fill}}lllr}
\hline
\hline
$L$ & $N_T$ & $N_{\rm samp}$ & $N_{\rm sw}$\\
\hline
4  & 16 & 10000 &  262144 \\
5  & 16 & 10000 &  262144 \\
6  & 16 & 10000 &  262144 \\
8  & 16 & 5000  & 1048576 \\
10 & 22 & 2500  & 8388608 \\ 
\hline
\hline
\end{tabular*}
\end{table}

\paragraph*{Disorder and Temperature Chaos}

In what follows we discuss how chaos can arise in spin glasses
using the early arguments presented in Refs.~\cite{mckay:82},
\cite{fisher:86}, \cite{bray:87}, and \cite{meanfield}. Within the
droplet theory framework \cite{bray:84,fisher:86}, the low-lying
excitations above the equilibrium state are obtained by flipping
compact connected clusters of spins called droplets.  A droplet of
size $\ell$ has a fractal surface of dimension $d_{\rm s} < d$, and
its excitation free energy $F > 0$ is distributed via $P_T(F,\ell) =
[ \gamma(T) \ell]^{-\theta} \rho [F / \gamma(T) \ell^{\theta}]$, where
$\rho(x)$ is a scaling function assumed to be nonzero at $x = 0$ and
which decays to zero for large $x$. The free-energy exponent $\theta$
is argued on general grounds to be such that $0 < \theta \leq d_{\rm
  s}/2$ and $\gamma(T)$ is the free-energy stiffness (which goes
to zero at $T_{\rm c}$). The droplet's entropy $S$ can be written
as $S = \sigma(T) \ell^{d_{\rm s}/2}$, where $\sigma$ is the
entropy stiffness.  Temperature chaos appears if the free energy
of a droplet changes its sign when the temperature is modified. As
noted in Refs.~\cite{fisher:86} and \cite{bray:87}, the length scale
at which this happens can be estimated by noting that the energy of
a droplet does not change much with temperature.  Therefore, if one
considers a droplet at temperature $T_1$ with free energy $F(T_1)$,
then at temperature $T_2>T_1$
\be 
F(T_2) \approx F(T_1) + T_1 S(T_1) - T_2 S(T_2) .  
\ee
Because for typical droplets $F(T_1)=\gamma(T_1) \ell^{\theta}$ and
$S = \sigma(T) \ell^{d_{\rm s}/2}$, the free energy excitation of such
droplets becomes generally negative at temperature $T_2$ (so that the droplet
has to be flipped) for length scales larger than the 
{\em chaotic length}~\cite{fisher:86,bray:87} defined as
\be 
\ell_{{\rm c}} = \bigg( \frac
{\gamma(T_1)} {T_2 ~\sigma(T_2) - T_1 ~\sigma(T_1)} \bigg)^{1/{\zeta}} \;\; \text{with}
\;\;\; \zeta=\frac{d_{\rm s}}{2}-\theta .  
\label{len2}
\ee
Usually, small temperature changes are studied such that $S(T_1) \approx
S(T_2)$. Here however, we do not use this approximation and since
in the low temperature phase, when one can define droplets, the entropy is
proportional to $\sqrt{T}$ \cite{fisher:86,aspelmeier:02a}, we write 
\be
\ell_{\rm c}(T_1,T_2) \propto \left( T_2^{3/2} - T_1^{3/2} \right)^{-1/\zeta} .
\label{len}
\ee
Equation (\ref{len}) shows that when temperature is changed,
equilibrium configurations are changed on scales greater than $\ell
_{{\rm c}}$. Notice that by keeping the temperature dependence of the
entropy, we obtain a slightly different scaling than usually considered
\cite{fisher:86,bray:87}, where a factor $\Delta T$ appears instead of
$T_2^{3/2}-T_1^{3/2}$. While this makes no difference for small $\Delta
T$ (which is the case in all simulations performed so far), this
can be significant for temperature differences larger than the ones
considered in this work.  Similar arguments can also be applied to the
case of a random perturbation in the disorder \cite{fisher:86,bray:87}
-- see for instance Ref.~\cite{krzakala:05} -- where one obtains
\be \ell_{\rm c}(\Delta J)
\propto  \Delta J^{-1/\zeta} .  
\ee 
Considering system-size
excitations, these arguments thus suggest that the chaoticity parameters
defined in Eqs.~(\ref{parameter1}) and (\ref{parameter2}) have the following
scaling behavior 
\bea
Q_{T_1,T_2}(L,T_1,T_2) &=& F_T[L/ \ell_{\rm c}(T_1,T_2)], \label{scaling} \\ \nonumber
Q_{\Delta J}(L,\Delta J) &=& F_J[L/ \ell_{\rm c}(\Delta J)] ,
\eea
where $F(x)$ is a function with $F(0) = 1$ that decays at large
$x$. In what follows we test the aforementioned scaling relations via
Monte Carlo simulations.

\paragraph*{Numerical results}

\begin{figure}
\includegraphics[width=0.8\columnwidth]{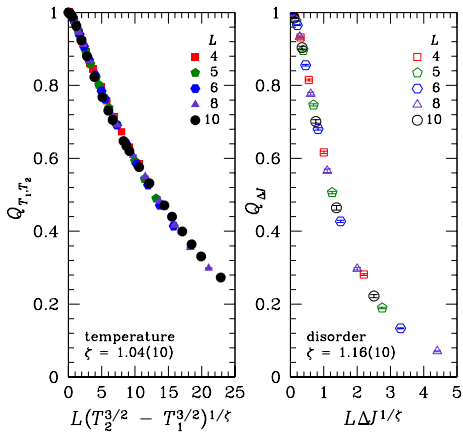}
\vspace*{-1.1cm}
\caption{(Color online)
Scaling plot of the chaoticity parameter for temperature chaos (left panel,
$T_1 = 0.20$) and disorder chaos (right panel, $T=0.2$) 
using $L/\ell_{\rm c}$ as a scaling variable.
}
\label{fig:chaos_scaling}
\end{figure}

\begin{figure}
\includegraphics[width=0.8\columnwidth]{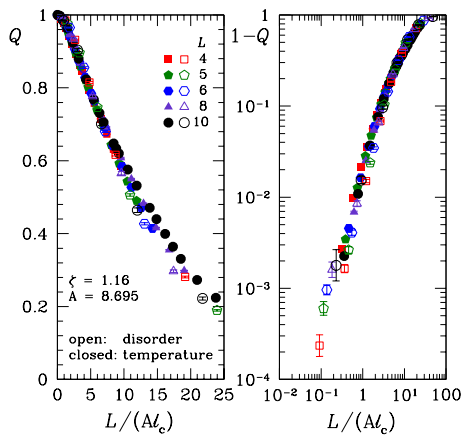}
\vspace*{-1.1cm}
\caption{(Color online)
Scaling plot of {\em both} disorder {\em and} temperature chaos. Left:
Chaoticity parameter $Q$ as a function of $L/\ell_{\rm c}$.  Scaling
performed with $\zeta=1.16$. Note that $\ell_{\rm c}(\Delta T) \approx A 
\ell_{\rm c}(\Delta J)$ with $A \approx 8.7$, i.e., 
disorder chaos appears at considerably
shorter length scales than temperature chaos. Right: Log-log plot of 
the same data, plotted as $1 - Q$, to illustrate the quality of the scaling.  }
\label{fig:both}
\end{figure}

The behavior of the chaoticity parameter for disorder chaos at zero
temperature has recently been studied in Ref.~\cite{krzakala:05}
(see also Ref.~\cite{rieger:96}) and perfectly satisfies the
predictions of the droplet model. We thus concentrate here on
finite temperatures where the scaling relations were only tested in
the simulations of Refs.~\cite{neynifle:97}, \cite{neynifle:98},
and \cite{sasaki:05} in two and four space dimensions. We have
performed low-temperature Monte Carlo simulations of the 3D Ising
spin glass (see Table \ref{tab:simparams}). As can be observed in
Fig.~\ref{fig:chaos_scaling}, scaling the data for disorder and
temperature chaos according to Eqs.~(\ref{scaling}) works extremely
well.  The best scaling collapse determined by a nonlinear minimization
routine \cite{katzgraber:06} yields $\zeta \approx 1.04$ for
temperature chaos and $\zeta \approx 1.16$ for disorder chaos, which
is in rather good agreement with the accepted value $\zeta \approx
1.1$ from $d_{\rm s} \approx 2.6$ \cite{palassini:00,katzgraber:01}
and $\theta \approx 0.2$ \cite{bray:84,mcmillan:84,hartmann:99};
see Eq.~(\ref{len2}). Notice that we have a good scaling of the
data even when $T_2$ is larger than $T_{\rm c}$.  We have tested the
temperature-dependence of the exponent $\zeta$ and find that for $T
\le 0.5$ its value is practically independent of temperature, i.e.,
at $T = 0.20$ we are probing the low-temperature regime.

Renormalization group arguments also suggest that the
temperature and disorder chaos effects are deeply related
and characterized by the same universal scaling function
\cite{fisher:86,bray:87,sales:02,silveira:04,doussal:05} so that
only nonuniversal prefactors differ.  In Fig.~\ref{fig:both},
we thus superimpose the data for both perturbations presented in
Fig.~\ref{fig:chaos_scaling} by rescaling $\ell_{\rm c}$.  Using
$\zeta=1.16$ for both perturbations and multiplying $\ell_{\rm c}
(\Delta J)$ by a factor $A \approx 8.7$, we obtain a rather good
superposition in the low-temperature region, and we conclude that our
data are thus compatible with the {\it equality} of the two scaling
functions. From this $\ell$-scale renormalization we also conclude that
the length scale at which temperature chaos appears is approximatively
ten times larger than the length scale needed for disorder chaos to
appear, as has also been found in four space dimensions by Sasaki {\em
et al}., see Ref.~\cite{sasaki:05}.  Therefore temperature chaos is
harder to probe than disorder chaos, as has already discussed within
a Migdal-Kadanoff approach on spin glasses \cite{aspelmeier:02a}.

The superposition of the scaling functions shows deviations at larger
temperatures. This is not surprising as assumptions made when deriving
the scaling function are not valid for high $T$. For example, the
$\sqrt{T}$ dependence of the entropy or the very existence of droplets
is only valid for $T \le T_c$ \cite{nifle:92}.  We thus believe that
to perform a definitive test of universality of the scaling functions,
very large system sizes at low temperatures with small temperature
changes should be used. 

We also study the behavior of the scaling function.  It decays as
$(\ell_{\rm c}/L)^{d/2}$ for strong chaos (when $\ell_{\rm c}/L \le
1$) \cite{krzakala:05}.  However, in the limit $\ell_{\rm c}/L \ge
1$, we obtain $1-Q(x)\propto x^{3 \zeta/2}$ which differs from the
$1-Q(x) \propto x^{\zeta}$ behavior found at zero temperature in
Ref.~\cite{krzakala:05}.  This shows that the scaling function can
have a more subtle behavior than what is naively expected from simple
domain-wall arguments \cite{sasaki:05,sheffler:03}. Note that the
results remain unchanged if one takes the disorder average of the
different overlaps in Eq.~(\ref{parameter1}) {\em independently},
as done in Eq.~(11) of Ref.~\cite{neynifle:98}.

\begin{figure}
\includegraphics[width=0.8\columnwidth]{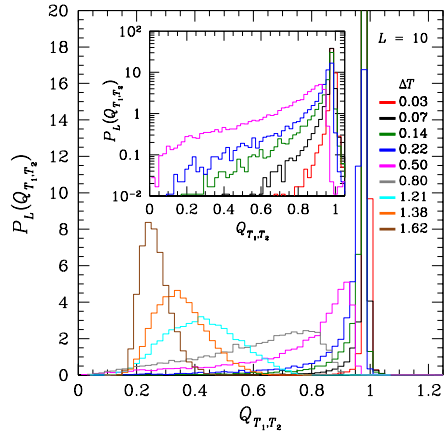}
\vspace*{-1.1cm}
\caption{(Color online)
Distribution of the chaoticity parameter over different realization of the
disorder for $L = 10$ with $T_1=0.2$ and 
$T_2=T_1+\Delta T$ \cite{temperatures}. In the case of
small $\Delta T$ rare samples have very low values of $Q$ while
most of them remain unchanged ($Q = 1$). The inset shows a linear-log plot of
the data for the smallest values of $\Delta T$ for which 
$T_1<T_2< 2 T_{\rm c}/3$. For all $T_2$ the distribution
has a single peak, unlike for the random-energy models with entropic
fluctuations \cite{krzakala:02,kurkova:03}.
}
\label{fig:distr}
\end{figure}

Finally, to better illustrate the mechanism of chaos, we study
the distribution of the chaoticity parameter over the disorder for
temperature chaos, i.e., we compute the chaoticity parameter $Q$
as defined in Eq.~(\ref{parameter1}) without the disorder average,
and bin the data for different choices of the disorder to compute the
distribution $P_{L}(Q_{T_1,T_2})$.  According to the droplet model, in
the weak chaos regime (where $\ell_{\rm c} > L$), temperature chaos can
manifest itself even on small length scales, but only for rare regions
of space \cite{sales:02}.  This means that even for small $\Delta
T$, when Q is very close to unity, the distribution is broad and
rare samples with lower values are expected.  Figure \ref{fig:distr}
shows the distribution $P_{L}(Q_{T_1,T_2})$ for $L=10$ for $T_1 = 0.2$
and different $T_2 = T_1 + \Delta T$ \cite{temperatures}.  Even for
modest $\Delta T$, rare but large changes are clearly observed. This
illustrates the weak chaos scenario presented in Ref.~\cite{sales:02}:
temperature chaos (at least in the weak regime) is not due to moderate
changes in all samples, but rather due to larger changes in a few
rare samples.

\paragraph*{Conclusions} We have studied numerically disorder
and temperature chaos in 3D Ising spin glasses and show that both
disorder as well as temperature chaos are well described within a
scaling/droplet description.  In particular, we find that the scaling
variables have to be modified as done in Eq.~(\ref{len}) when the
difference in temperature is large. In addition, we show that the weak
chaos regime is dominated by rare events where system-size droplets are
flipped.  This has direct experimental implications because the weak
chaos regime has been argued to account in a {\em quantitative} way
for the memory and rejuvenation effects \cite{jonsson-ea:04}.  Finally,
we show that temperature and disorder chaos might be described
by similar scaling functions in the low-temperature regime, thus
providing compelling evidence for the presence of a chaotic temperature
dependence in spin glasses. This has also recently been proven for
mean-field systems \cite{rizzo:06,yoshino:06}. This mechanism is also
responsible for step-size responses that could in principle be observed
experimentally in mesoscopic systems \cite{rizzo:06,yoshino:06}.
Nevertheless, this behavior might change for larger system sizes
and thus we propose to revisit the problem with better models
\cite{katzgraber:05d}.  Our findings will help interpret experiments
on rejuvenation and memory effects in spin-glasses and other materials.

\begin{acknowledgments}
We thank A.~Billoire, J.-P.~Bouchaud, T.~J\"org, M.~Sasaki,
H.~Yoshino, A.~P.~Young, and L.~Zdeborov\`a for discussions. The simulations 
have been performed on the Hreidar and Gonzales clusters at ETH Z\"urich.
\end{acknowledgments}

\vspace*{-0.7cm}

\bibliography{refs,comments}

\begin{thebibliography}{43}
\bibitem[{mea()}]{meanfield}
\bibinfo{note}{Note similar arguments can be used for mean field systems
  \cite{krzakala:02,kurkova:03,rizzo:06} leading to similar conclusions.}
\end{thebibliography}

\end{document}